\documentclass[amsmath,showpacs]{revtex4}
\begin{document}

\title{Exact transmission moments in one-dimensional weak localization and single-parameter scaling}
\author{Jean Heinrichs}
\email{J.Heinrichs@ulg.ac.be}
\affiliation{Institut de Physique, B5, Universit\'{e} de Li\`{e}ge, Sart
Tilman, B-4000 Li\`{e}ge, Belgium}

\begin{abstract}

We obtain for the first time the expressions for the mean and the variance of the transmission coefficient for an Anderson 
chain in the weak localization regime, using exact expansions of the complex transmission- and reflection coefficients to
fourth order in the weakly disordered site energies.  These results confirm the validity of single-parameter scaling theory in
a domain where the higher transmission cumulants may be neglected.  We compare our results with earlier results for
transmission cumulants in the weak localization domain based on the phase randomization hypothesis.
\end{abstract}


\maketitle

\section{Introduction}

The validity of the single-parameter scaling hypothesis in the scaling theory of localization of Abrahams {\it et al.} \cite{1}
has been the object of numerous discussions during the decade following its publication in 1979.  This scaling theory assumes
that the scaling of a typical conductance $g$ as a function of size $L$ of a $d$-dimensional disordered sample is a function of
only one parameter, namely the typical conductance itself.

Soon after the appearance of Ref.\cite{1} it became clear that one should consider the scaling of the full probability 
distribution of conductance (or the distribution of a related scaling variable) rather than simply that of a typical
conductance
\cite{2}. Single-parameter scaling then requires that the distribution of the chosen variable should scale to a universal form
depending only on one parameter.  A simple example of such a statistical distribution is the familiar normal distribution in
which the large scale dispersion $\sigma^2$ decays to zero via the law of large numbers relating it to the growth rate of the
mean value, the only parameter of this distribution.

Many aspects of scaling of distributions of transport related quantities such as the resistance $\rho=\frac{1}{g}$ have been
discussed some  years ago by Shapiro and coworkers
\cite{3} and in a more recent review \cite{4}.

Until recently  analytical discussions of single-parameter scaling have invariably relied on the so-called phase randomization 
hypothesis
\cite{2}.  This hypothesis assumes the existence of a microscopic length scale, shorter than the localization length, over
which the phases of the complex reflection- and transmission coefficients of a disordered sample are randomized in such a way
that their distribution becomes uniform.

Let us briefly recall the scaling results for a 1D-disordered chain of length $L$.  In this case a convenient scaling 
variable related to the conductance is the finite scale Lyapunov exponent\cite{2}

\begin{align}\label{eq1}
\bar\gamma_L&=\frac{1}{2L}\ln\left(1+\frac{1}{g}\right)\nonumber\\
&=-\frac{1}{2L}\ln t_L
\quad ,
\end{align}

\noindent
where the second line follows from using the Landauer formula,$\rho=\frac{1}{g}=\frac{r_L}{t_L}$ , which relates the resistance
to the transmission ($t_L$) and  reflection $(r_L)$ coefficients of the chain.  The advantage of the scaling variable
$\bar\gamma_L$ is that under the assumption of complete phase randomization it obeys a normal distribution
\cite{3,4} of the type alluded to above, in the strong localization regime, $L>>L_c$ (where
$L_c^{-1}\equiv\gamma=\lim_{L\rightarrow\infty}\gamma_L$ is the inverse localization length), for weak disorder.  The
distribution is centered at the asymptotic value $L_c^{-1}$, with a dispersion 

\begin{equation}\label{eq2}
\sigma^2=\text{var}\;\gamma_L=\frac{\gamma}{L}
\quad ,
\end{equation}

\noindent
i.e. it describes single-parameter scaling in terms of $\gamma=L_c^{-1}$.  The Eq. \eqref{eq2}coincides with the general large
number law property  for the validity of single-parameter scaling (SPS) for $\bar\gamma_L$ proposed recently by Deych {\it et
al.} for the general case where phase randomization is not hypothesized. In their work \cite{5} Deych {\it et al.} studied the
condition
\eqref{eq2} analytically for an Anderson chain with a Cauchy distribution of site energies.  This has led them to identify a
new characteristic length beyond which single-parameter scaling holds true in their system \cite{5}.  On the other hand,
under the random phase assumption one-parameter scaling also exists in the weak localization (quasi metallic) regime for weak
disorder \cite{3,4}, where  $L<<L_c$.  In this case the resistance $\rho<<1 (\rho\simeq 2L\bar\gamma)$ follows an exponential
(Poisson) distribution
\cite{3,4} depending on the sole parameter $L^{-1}_c$.  The condition for one-parameter scaling analogous to \eqref{eq2} then
reads (see also Sect. III)

\begin{equation}\label{eq3}
\sigma^2=\langle\bar\gamma_L\rangle^2
\quad ,
\end{equation}

\noindent
where $\langle\ldots\rangle$ denotes averaging over the disorder.  The Eq. \eqref{eq3} implies one-parameter scaling in a
domain where the higher  cumulants of the resistance distribution are subdominant.  We also note that \eqref{eq3} is exactly
verified (with $\langle\bar\gamma_L\rangle^2=\gamma^2$) by the exponential distribution of resistance obtained in the
framework of the random phase assumption for $L<<L_c$. 

The purpose of this paper is to analyse the validity of one-parameter
scaling theory in the weak localization regime, by deriving exact analytical results for the mean value and variance of the
transmission coefficient $t_L$ for a weakly disordered Anderson chain, without using phase randomization.  The resulting
expressions for the variance $\sigma^2$ and the mean value $\langle\bar\gamma_L\rangle$ of the variable $\bar\gamma_L$ are
shwon to obey approximately (for $L>>a$, with $a$ the lattice parameter) a universal relation analogous to
\eqref{eq3}.  Here the mean  $\langle\bar\gamma_L\rangle$, which turns out to depend on $L$ is completely defined in terms of a
fixed localization length. The complex transmission coefficient $T_L(t_L=|T_L|^2$ and reflection coefficient $R_L
(r_L=|R_L|^2)$ are obtained in Sect. II by iterating the exact coupled recursion relations obeyed by these quantities
\cite{6,7}, for weak disorder in the regime $L<<L_c$.  These results are used for obtaining explicit forms of the first two
moments of the transmission coefficient $t_L$ and of the corresponding Lyapunov exponent $\bar\gamma_L$.  These moments are
discussed in connection with the validity of one-parameter scaling theory in Sect. III, where they are also compared with
previous results based on the random phase assumption.

\section{Transmission in the weak localization regime}

Consider a disordered linear chain of length $L=Na$ composed of $N$ disordered one-orbital sites of spacing $a=1$ described by
the  Schr\"{o}dinger equation

\begin{equation}\label{eq4}
\varphi_{n+1}+\varphi_{n-1}+\varepsilon_n\varphi_n=E\varphi_n
\quad ,
\end{equation}

\noindent
where $\varphi_m$ is the wavefunction amplitude and $\varepsilon_m$ the random energy of the site $m$ measured (like $E$
also) in units of the constant hopping rate between nearest-neighbour sites.  We assume the
$\varepsilon_m$ to be identically distributed  independent gaussian variables with mean zero and correlation

\begin{equation}\label{eq5}
\langle\varepsilon_m\;\varepsilon_n\rangle=\varepsilon^2_0\delta_{m,n}
\quad .
\end{equation}

\noindent
The usual Anderson model of uniformly distributed site energies between $-W$ and $W$ correspond to
$\varepsilon^2_0=\frac{W^2}{12}$.  The disordered chain is  connected to electron reservoirs by semi-infinite non-disordered
leads defined by sites $m>N$ and $m<1$, respectively.   As indicated in Sect. I, we whish to calculate the transmission
coefficient $t_N=|T_N|^2$ and to study its statistical moments in the weak  localization regime $L<<L_c$, for weak disorder. 
The inverse localization length is defined by

\begin{equation}\label{eq6}
\gamma=\frac{1}{L_c}=-\lim_{N\rightarrow\infty}\frac{1}{2N}\ln t_N
\quad .
\end{equation}

\noindent
The complex transmission- and reflection amplitudes for a plane wave incident from the right ($n>N$) with wavenumber-$k$ and
energy

\begin{equation}\label{eq7}
E=2\cos k,\quad 0\leq k\leq \pi
\quad ,
\end{equation}

\noindent
obey the exact coupled recursion relations \cite{6,7}

\begin{equation}\label{eq8}
T_N=\frac{e^{ik}T_{N-1}}{1-i\nu_N(1+e^{2ik}R_{N-1})}
\quad ,
\end{equation}

\begin{equation}\label{eq9}
R_N=\frac{e^{2ik}R_{N-1}+i\nu_N(1+e^{2ik}R_{N-1})}{1-i\nu_N(1+e^{2ik}R_{N-1})}
\quad ,
\end{equation}

\begin{equation}\label{eq10}
\nu_N=\frac{\varepsilon_N}{2\sin k}
\quad ,
\end{equation}

\noindent
which relate the amplitudes $T_N(R_N)$ for the chain of $N$ disordered sites to those for a chain of $N-1$ sites, with the
initial  conditions $T_0=1$ and $R_0=0$.  The choice of wavenumbers in \eqref{eq7} is such that the leadsÕ eigenfunctions 
$\varphi_n\sim e^{\pm ikn}$ correspond to Bloch waves traveling from left to right and from right to left, respectively. The
solution of
\eqref{eq9} may be readily obtained in terms of the reflection amplitudes for chains whose lengths are comprised between $1$
and
$N-1$:

\begin{equation}\label{eq11}
T_N=e^{ikN}\prod^N_{n=1}\frac{1}{1-i\nu_n(1+e^{2ik}R_{N-1})}
\end{equation}

For weak disorder we expand \eqref{eq11} to successive orders in the site energies $\varepsilon_n$.  As will be readily seen
below the validity  of this perturbation expansion is restricted typically to chain lengths $L<<L_c$ (the weak localization
regime),  where $L_c$ is the localization length for weak disorder first derived by Thouless \cite{8}.  The study of the
dispersion of the transmission coefficient requires to include terms up to fourth order in the expansion of \eqref{eq11}.  We
thus obtain

\begin{align}\label{eq12}
T_N&=e^{ikN}[1+\sum^N_{m=1}(P_m-1)
+\sum^N_{\begin{matrix}m,n=1\\m\neq n\end{matrix}}(P_m-1)(P_n-1)\nonumber\\
&+\sum_{\begin{matrix}m,n,p\\m\neq n\neq p\neq m\end{matrix}}(P_m-1)(P_n-1)(P_p-1)
+\sum_{\begin{matrix}m,n,p,q\\m\neq n\neq p\neq q\\n\neq q\neq  m\neq
p\end{matrix}}(P_m-1)(P_n-1)(P_p-1)(P_q-1)\quad ,\nonumber\\ 
P_j&=\frac{1}{1+x_j}\;,\; x_j=-i\nu_j(1+e^{2ik}R_{j-1})
\quad.
\end{align}

\noindent
In each one of the successive terms in \eqref{eq12}, we then retain contributions up to fourth order in the variables $\nu_j$,
via the  expansion of the $P_j$ in power of $x_j$ and corresponding expansions of $R_j$ in successive order contributions in
the reflection amplitudes,

\begin{equation}\label{eq13}
R_n=\sum^4_{q=1} R_n^{(q)}\quad .
\end{equation}

\noindent
The contributions $R_n^{(q)}$ at successive orders $q$ are determined by the perturbation equations obtained by expanding the
recursion  equation analogous to \eqref{eq9} relating $R_n$ and $R_{n-1}$ for systems of length $1\leq n\leq N-1$ and 
$n-1$: 

\begin{subequations}
\begin{align}\label{eq14}
R_n^{(1)}&=e^{2ik}R_{n-1}^{(1)}+i\nu_n\quad ,\\
R_n^{(2)}&=e^{2ik}R_{n-1}^{(2)}+2i\nu_n e^{2ik}R^{(1)}_{n-1}-\nu^2_n\quad ,\\
R_n^{(3)}&=e^{2ik}R_{n-1}^{(3)}+2i\nu_n e^{2ik}R^{(2)}_{n-1}+i\nu_n e^{4ik} R_{n-1}^{(1)2}-2\nu^2_n e^{2ik}
R^{(1)}_{n-1}-i\nu^3_n\quad,
\end{align}
\end{subequations}

\noindent
whose solutions are

\begin{subequations}
\begin{align}\label{eq15}
R^{(1)}_n&=i\sum^n_{\rho=1}\nu_p e^{2ik(n-p)}\quad ,\\
R^{(2)}_n&=-\sum^n_{p=1}\nu_p e^{2ik(n-p)}\left[\nu_p+2\sum^{p-1}_{q=1}\nu_q e^{2ik(p-p)}\right]\quad ,\\
R^{(3)}_n&=-i\sum^n_{p=1}\nu_p e^{i2(n-p)}\left[\nu^2_p-2i\nu_p R_{p-1}^{(1)}e^{2ik}
-R_{p-1}^{(1)2}e^{4ik}-2R_{p-1}^{(2)}e^{2ik}\right]\quad .
\end{align}
\end{subequations}

\noindent
Next, by using (\ref{eq12}-\ref{eq13}) and (15.a-15.c), we express the transmission coefficient $t_N=|T_N|^2$ to fourth order
in the site  energies and, finally, we find its average $\langle t_N\rangle$ using \eqref{eq5} together with the factorization
property for independent gaussian variables,

\begin{equation}\label{eq16}
\langle \varepsilon_m\varepsilon_n\varepsilon_p\varepsilon_q\rangle=
\langle\varepsilon_m\varepsilon_n\rangle
\langle\varepsilon_p\varepsilon_q\rangle+
\langle\varepsilon_m\varepsilon_p\rangle
\langle\varepsilon_n\varepsilon_q\rangle+
\langle\varepsilon_m\varepsilon_q\rangle
\langle\varepsilon_n\varepsilon_p\rangle\quad ,
\end{equation}

\noindent
which is valid for arbitrary indices $m,n,p,q$ different or not \cite{9}.

These straightforward but rather tedious calculations yield the final result

\begin{equation}\label{eq17}
\langle t_N\rangle=1-\frac{2N}{L_c}-\frac{2N^2}{L^2_c}\left[5-\frac{1}{N}-\frac{2}{N^2}\frac{\sin^2 kN}{\sin^2 k}
\left(3+\frac{\cos^2 kN}{\cos^2 k}\right)\right]+O (\varepsilon^6_0)\quad ,
\end{equation}

\noindent
which is exact to quadratic order in $\varepsilon_0^2$.  Here

\begin{equation}\label{eq18}
\frac{1}{L_c}=\frac{\varepsilon^2_0}{8\sin^2 k}=\frac{\varepsilon^2_0}{2(4-E^2)}\quad ,
\end{equation}

\noindent
is Thouless expression \cite{8} for the localization length, which is valid to lowest order in the weak disorder.  The 
perturbation expression \eqref{eq17} shows that besides the smallness of $\varepsilon^2_0$ its validity requires typically
$N<<L_c$, which implies that the eigenstates are delocalized on the scale of the chain length.  We further note that while the
lowest correction in $\langle t_N\rangle$ for weak disorder is of order $\frac{N}{L_c}$, the next higher contribution
proportional to $\varepsilon_0^4$ includes three types of terms respectively proportional to $\left(\frac{N}{L_c}\right)^2$,
$\left(\frac{N}{L_c}\right)\varepsilon_0^2$ and $\varepsilon_0^4$.  By a similar calculation of the second moment, $\langle
t_N^2\rangle$, we obtain

\begin{equation}\label{eq19}
\langle t^2_N\rangle=1-\frac{4N}{L_c}+\frac{4N^2}{L^2_c}
\left[2-\frac{2}{N}+\frac{1}{N^2}\frac{\sin^2 kN}{\sin^2 k}\left(4+\frac{2\cos^2 k N}{\cos^2 k}-\frac{\sin^2
kN}{2\sin^2 k}\right)\right] +O(\varepsilon^6_0)\quad .
\end{equation}

\noindent
The calculation of $\langle t_N^2\rangle$ has been greatly facilitated by making use of the expression of $t_N$
to quartic order and of the corresponding average \eqref{eq17}.  From \eqref{eq19} and \eqref{eq18} we finally get

\begin{equation}\label{eq20}
\text{var}\;t_N=\frac{12N^2}{L^2_c}\left[2-\frac{1}{N}-\frac{1}{3N^2}\frac{\sin^2 kN}{\sin^2 k}
\left(2+\frac{\sin^2 kN}{2\sin^2 k}\right)\right]+O (\varepsilon^6_0)\quad .
\end{equation}

The first and second moments and the variance of $\bar\gamma_L$ in \eqref{eq1} are easily found from \eqref{eq17} and
\eqref{eq19} by expanding  $\ln t_N$ in powers of the small quantity $1-t_N$.
Thus we obtain

\begin{equation}\label{eq21}
\langle\bar\gamma_N\rangle=-\frac{1}{2 N}\langle\ln t_N\rangle=\frac{1}{L_c}\left(1+\frac{4N}{L_c}
\left[3-\frac{1}{N}-\frac{1}{2N^2}\frac{\sin^2 kN}{\sin^2 k}\left(4+\frac{\cos^2 kN}{\cos^2
k}+\frac{\sin^2 kN}{4\sin 2 k}\right)\right]\right)+O (\varepsilon^6_0)\quad ,
\end{equation}

\begin{equation}\label{eq22}
\langle\bar\gamma_N^2\rangle=\frac{1}{L^2_c}\left[7-\frac{3}{N}-\frac{1}{N^2}\frac{\sin^2 kN}{\sin^2
k}\left(2+\frac{\sin^2 kN}{2\sin 2 k}\right)\right]+O (\varepsilon^6_0)\quad ,
\end{equation}

\begin{equation}\label{eq23}
\sigma^2=
\text{var}\;\bar\gamma_N=
\frac{3}{L^2_c}
\left[2-\frac{1}{N}-\frac{1}{N^2}\frac{\sin^2 kN}{\sin^2 k}
\left(2+\frac{\sin 2 kN}{2\sin 2 k}\right)\right]\quad .
\end{equation}

\noindent
By ignoring the terms of orders $\frac{1}{N}$ and $\frac{1}{N^2}$ relative to unity in the square bracketts of 
(\ref{eq21}-\ref{eq23}) it thus follows that

\begin{equation}\label{eq24}
\sigma^2=6\left(\langle\bar\gamma_N\rangle\right)^2+O (\varepsilon^6_0)\quad .
\end{equation}

\section{Discussion and concluding remarks}

The equation \eqref{eq24} is a universal relation expressing the variance of the scaling variable $\bar\gamma_N$ in terms of
the mean,  which is itself defined in terms of the Thouless localization length \eqref{eq18} alone, when terms of order
$\frac{1}{N}$ and $\frac{1}{N^2}$ are neglected relative to unity in \eqref{eq21}.  It is natural to expect that, within the
same approximation, similar universal forms in terms of $\langle\bar\gamma_N\rangle$ exist for the higher fluctuation cumulants
of
$\bar\gamma_N$.  Thus our work suggests that the exact distribution of $\bar\gamma_N$ (which does not rely on phase
randomization) in the weak localization regime obeys single-parameter scaling.  Note that
\eqref{eq24} differs by a factor of 6 from the universal relation \eqref{eq3} for the weak localization regime obtained under
the phase randomization assumption. We observe in passing that the mean of $\bar\gamma_N$ in \eqref{eq21} differs from the
Thouless inverse localization lengths
\eqref{eq18}.  This is because the inverse localization length is by definition, the non-random central limit of
$\bar\gamma_N$.  In order to obtain $L^{-1}_c$ from our analysis for $N<<L_c$ one has to let $|\varepsilon_0|\rightarrow 0$
before taking the large
$N$ limit.  In this limit only the leading order term of the weak disorder expansion of $\langle t_N\rangle$ in \eqref{eq17} is
relevant for finding  $L^{-1}_c$.  We believe that the reason why this term yields the correct (Thouless) localization length
in spite of the fact $\bar\gamma_N$ is not self-averaging in the domain $N<<L_c$ is because the non-self averaging is only
marginal i.e. $\lim_{N\rightarrow\infty}\lim_{|\varepsilon_0|\rightarrow 0}\frac{\sigma^2}{\langle\gamma_N\rangle^2}=$constant
in this case.

The aim of the present work has been to discuss the scaling of $\bar\gamma_L$ in the weak localization regime without using the
phase  randomization hypothesis.  In this context we now compare the above results for the transmission coefficient moments
with earlier results \cite{10} obtained from an invariant imbedding model \cite{11,12} analysed by assuming phase rendomization
\cite{10}.  The coupled invariant imbedding equations for the complex reflection $(R_N)$- and transmission coefficients $(T_N)$
\cite{11,12} have since been shown \cite{6} to correspond to the continuum limit of the Schr\"{o}dinger  equation for the
one-dimensional Anderson model for weak disorder.  This makes the comparison of the present results with those of
ref.\cite{10} the more relevant.  Under the random phase assumption we obtained the following exact results \cite{13} for the
transmission coefficient moments for $L<<L_c$

\begin{equation}\label{eq25}
\langle t^m_N\rangle=\left(1+2m\frac{N}{L_c}\right)^{-1}\;,\; m=0,1,2,3,\ldots \quad ,
\end{equation}

\noindent
which leads to the distribution of the transmission coefficient \cite{10}

\begin{equation}\label{eq26}
P_t (t_N)=\frac{L_c}{2N}t^{(L_c/2N-1)}\quad ,
\end{equation}

\noindent
which depends on the single scale parameter $L_c$.  Note that the exact moments \eqref{eq17} and \eqref{eq19} which do not rely
on the  random phase assumption reveal important deviations from Eq \eqref{eq25} at order  $\varepsilon_0^4$, even when the
terms in
$\frac{1}{N}$ and $\frac{1}{N^2}$ in the square bracketts are neglected \cite{14}.  The exact random phase expression for 
$\text{var}\; t_N$,

\begin{equation}\label{eq27}
\text{var}\;t_N=\frac{4N^2}{L^2_c\left(1+\frac{4N}{L_c}\right)\left(1+\frac{2N}{L_c}\right)^2}\quad ,
\end{equation}

\noindent
obtained from \eqref{eq25} differs also from the exact expression \eqref{eq20} to order $\frac{N^2}{L^2_c}$.  For
completenessÕ sake we finally list  the expressions for the first and second moments and the variance of $\bar\gamma_N$
obtained from expanding the logarithm of
\eqref{eq25} for $m=1$ and $m=2$ through order $\varepsilon_0^4$:

\begin{equation}\label{eq28}
\langle\bar\gamma_N\rangle=\frac{1}{L_c}\left(1-\frac{2N}{L_c}\right)\quad ,
\end{equation}

\begin{equation}\label{eq29}
\langle\bar\gamma^2_N\rangle=\frac{4}{L^2_c}\quad ,
\end{equation}

\begin{equation}\label{eq30}
\sigma^2=\left(\langle\bar\gamma_N\rangle\right)^2\quad ,
\end{equation}


\begin{thebibliography}{unsrt}
\bibitem{1} E. Abrahams, P.W. Anderson, D.C. Licciardello and T.V. Ramakrishnan, Phys. Rev. Lett. {\bf 42}, 673 (1979).
\bibitem{2} P.W. Anderson; D.J. Thouless, E. Abrahams and D.S. Fisher, Phys. Rev. B{\bf 22}, 3519 (1980).
\bibitem{3} B. Shapiro, Phil. Mag. B{\bf 56}, 1031 (1987); A. Cohen, Y. Roth and B. Shapiro, Phys. Rev. B{\bf 38}, 12125
(1988).
\bibitem{4}M. Janssen, Phys. Rep. {\bf 295}, 1 (1998).
\bibitem{5} L.I. Deych; A.A. Lisansky and B.L. Altshuler, Phys. Rev. Lett. {\bf 84}, 2678 (2000); Phys. Rev. B{\bf 64}, 224202
(2001).
\bibitem{6} J. Heinrichs, Phys. Rev. B{\bf 65}, 075112 (2002).
\bibitem{7} C. Barnes and J.M. Luck, J. Phys. A{\bf 23}, 1717 (1990).
\bibitem{8} D.J. Thouless, in Ill-Condensed Matter, edited by R. Balian, R. Maynard and G. Toulouse (North-Holland, Amsterdam,
1979). 
\bibitem{9} See e.g. N.G. Van Kampen, Stochastic Processes in Physics and Chemistry (North-Holland, Amsterdam, 1981).
\bibitem{10} J. Heinrichs, J. Phys.: Condensed Matter, {\bf 1}, 2651 (1989).
\bibitem{11} R. Bellman and C.M. Wing, An Introduction to Invariant Imbedding (Wiley, New-York, 1976).
\bibitem{12} For a review of invariant imbedding and applications, see R. Rammal and B. Dou\c{c}ot, J. Phys. (Paris) {\bf 48},
509 (1987).
\bibitem{13} We recall that the derivation of Ref. \cite{6} leads to a natural definition of the localization length in the
framework of the invariant imbedding model: it is simply given by the continuum limit of the localization length
\eqref{eq18}.  This is twice the localization length which was used in ref. \cite{10} (which corresponded to a different
definition of the mean free path in the classical resistivity
\cite{2} in terms of parameters of the invariant imbedding model) and leads to the coefficient 2 on the r.h.s. of
\eqref{eq25}. 
\bibitem{14} Using the Landauer formula, $\rho=t^{-1}-1$, the distribution \eqref{eq26} was shown in \cite{10} to yield the
familiar exponential distribution of resistance for $\rho<<1\;(L<<L_c)$\cite{3}.  Actually, however, this derivation yields the
slightly more accurate expression $P_N(\rho)=l^{-1}\exp[-(l^{-1}+1)\rho], \; \ell=2 \frac{N}{L_c}$.  The improved expression
turns out to be necessary for recovering
\eqref{eq26} and
\eqref{eq25}, starting from the distribution of resistance.

\end{thebibliography}
\end{document}